\documentstyle[psfig,preprint,aps]{revtex}
\begin{document}
\draft
\title{Fingerprints of entangled states in reactions with rare isotopes}
\author{ C.A. Bertulani $^{\dag}$\footnotetext{$^{\dag}$
E-mail: bertulani@nscl.msu.edu} }
\address{
Department of Physics and Astronomy and National Superconducting Cyclotron
Laboratory, Michigan State University, East Lansing, MI 48824-1321 - USA}
\date{\today}
%\date{February 20, 2002}
\maketitle

\begin{abstract}
We study the presence of entangled states of nucleon pairs from
nuclear decays and in reactions with exotic nuclei, e.g.,
$^{11}$Li, or $^6$He. It is shown that the fingerprints of
entangled states in these subsystems are visible in correlation
measurements and can be accessed with present experimental
techniques. This shows that not only atomic and optical systems,
but also nuclear systems serve as important tools to obtain
dichotomic outcomes for tests of the Einstein-Podolsky-Rosen
paradox.

\end{abstract}

\smallskip
\section{Introduction}
\label{sec:intro}

In recent years, there has been much interest in the physics of
unstable nuclei. Nuclei far from the valley of nuclear stability,
with lifetime of a few miliseconds, play an important role in
cosmology and astrophysics, as well as in the traditional
applications of nuclear physics. Some of these nuclei possess
rather unexpected properties, as for example, the two-proton
radioactivity \cite{Gio02}, nuclear halos, large deformations,
etc. Many theoretical models for these nuclei invoke subtle
aspects of basic quantum mechanics. A nuclear halo, for example,
is thought to be a simple manifestation of quantum tunneling of
loosely bound states. To gain insight into these features, typical
experiments involve the measurement of momentum distributions of
the fragments, knock-out and stripping reactions, Coulomb
excitation, etc.

Rare nuclear isotopes can also be used to study counterintuitive aspects of
quantum mechanics. For example, two-proton decay in s-wave states could
be used for a test of quantum mechanics versus local realism
by means of Bell's inequalities \cite{Bel64}. Since the final state of the two protons
can be found in a singlet state, their wave function is spin-entangled. The identification
of the spins of  the proton in two detectors separated far away would be useful to
test the  Einstein-Podolski-Rosen (EPR) paradox \cite{EPR35}. In fact, these tests should be performed
in different and complementary branches of physics to avoid the loopholes encountered
in photon experiments.
Correlation experiments with low-energy proton-proton scattering have already been used
for studies on the EPR paradox. Two-proton radioactivity would also qualify for this purpose.

Simple quantum entanglements are also visible in peripheral
reactions with rare isotopes \cite{Mar01}. We will consider some
examples here. To start with, we observe that in Coulomb
dissociation experiments of nuclear collisions at intermediate
energies ($\gtrsim 50$ MeV per nucleon), the exchanged photon is
almost real, thus the momentum and energy transferred obey the
relation $p\simeq E/c$. A single nucleon cannot absorb this photon
(the photon momentum is too small, $p\simeq 0$ for $E\simeq 1$
MeV). In order to preserve the energy-momentum condition the
photon has to be absorbed by more than one nucleon. The energy can
be shared by two nucleons which fly apart in opposite directions.
The same happens in nuclear interactions in peripheral collisions,
when little energy is transferred to the nucleus. At forward
scattering angles the momentum transferred is $|{\bf p-p'}|\simeq
p-p'\cos \theta\simeq p-p'\simeq E/v$, where $\bf p$ ($\bf p'$) is
the initial (final) center of mass energy, $\theta$ is the
scattering angle, and $v$ the projectile velocity. For collisions
at intermediate energies, $v\simeq c$, and the same argument as
for the Coulomb dissociation applies. Experiments which explore
this feature have been performed. For example, Coulomb
dissociation of $^{11}$Li projectiles (a rare isotope) into a
$^9$Li nucleus and two neutrons have been done \cite{Iek93}. Since
neither $^{10}$Li, nor the di-neutron system are bound, the
neutron-neutron correlation in the bound state of $^{11}$Li is the
key factor to allow the mere existence of this nucleus.  The
experiments have concentrated their efforts in understanding this
particular aspect of the $^9$Li + n + n system.

The singlet and triplet
configurations of the neutron-neutron system in $^{11}$Li have different contributions to the
total nuclear wavefunctions and a recent experiment \cite{Sim99} has been able to access their relative
weights. However, some assumptions on the reaction mechanism were necessary to interpret the results.
It is also not
clear if different spin configurations in the final state could change the momentum
distributions of the fragments. In fact, very little is known about final state configurations
in reactions involving unstable nuclei.

In this article we show that one can disentangle the contributions
of singlet and triplet spin final states of a  two nucleon system
in reactions with rare isotopes by measuring momentum
correlations. This is a useful result as a direct measurement of
the spin orientations of each nucleon is by far more complicated
in such conditions. The application of the method is very general
as it only relies on measured quantities, independent of models
for the reaction mechanism. These are the widths and differential
cross sections of momentum distributions, which have been obtained
experimentally for several reactions involving unstable nuclei.

The correlated (C) and uncorrelated (U) measurements of nucleons 1 and 2, with
relative momentum $\Delta p = |{\bf p}_1-
{\bf p}_2|$ are defined by
\begin{equation}
C(\Delta p)=\int P\left({\bf p}_1; {\bf p}_1 + \Delta {\bf p}\right) d{\bf p}_1 d\Omega_p \ ,
\ \ \ \ \ \ \ \
U(\Delta  p)={1\over N}
\int P\left({\bf p}_1\right) P\left({\bf p}_1 + \Delta {\bf p}\right) d{\bf p}_1
d\Omega_p \ ,
\label{CU}
\end{equation}
where $P\left({\bf p}_1; {\bf p}_2 \right)$ is the probability to measure
a nucleon having momentum ${\bf p}_1$ in coincidence with the measurement of the other nucleon
having momentum ${\bf p}_2$.  The integration in $\Omega_p$ is over all orientations of
$\Delta {\bf p}$.
$P\left({\bf p}\right)=\int P\left({\bf p}; {\bf p}' \right)d{\bf p}'$ is the  probability
to measure the momentum ${\bf p}$ for one of the nucleons,  irrespective of what the momentum of
the other nucleon is. $N$ is the total number of particle measured, i.e., $N=\int
P\left({\bf p}\right) d{\bf p}$.

The correlation function is defined as
\begin{equation}
R\left( \Delta  p\right) ={C(\Delta  p) \over U(\Delta  p)} -1 .
\label{R}
\end{equation}

One can describe each nucleon in the final state
by Gaussian wavepackets:
\begin{equation}
\psi_i = \left( 2\pi \sigma_n^2\right)^{-3/4} \exp \left\{
-{
\left(
{\bf p}_i- ( {\bf P}_c \pm {\bf p}_n )/2
\right)^2
\over 4\sigma_n^2
}
\right\} ,
\end{equation}
where  the average recoil momentum of the core is ${\bf P}_c$, and the average
momentum of the pair at their center of mass is ${\bf p}_n$.
The momentum spread of the nucleon wavefunctions in the final state, $\sigma_n$, depends on the
reaction mechanism for the specific reaction studied.
E.g., for the breakup of $^{11}$Li projectiles
at $E_{lab} \simeq 300$ MeV/nucleon, the neutrons acquire
a momentum spread of the order of $\sigma_n \simeq 20$ MeV/c.

The spins of low energy nucleons imply that they can be arranged into singlet ($S=0$) and triplet
($S=1$) states. The spatial part of their wave function is symmetrized accordingly:
\begin{equation}
\Psi^{(\pm)}\left( {\bf p}_1,{\bf p}_2;{\bf P}_c, {\bf p}_n\right)
 = A^{\pm} \left\{ \psi_1({\bf p}_1;{\bf P}_c, {\bf p}_n)
\psi_2({\bf p}_2;{\bf P}_c, {\bf p}_n)\pm \psi_1({\bf p}_2;{\bf P}_c, {\bf p}_n)
\psi_2({\bf p}_1;{\bf P}_c, {\bf p}_n)
\right\}.
\end{equation}
This wavefunction is entangled and cannot be factorized into individual particle wavefunctions.
It will lead unavoidably to interferences which are visible if one observes the complete final
state. This non-local entanglement is an example of the EPR paradox.
The plus (minus) sign refers to $S=0$ ($S=1$) states, $A^{\pm} = \left[
2(1\pm {\cal O}^2)\right]^{-1/2}$,
where the overlap integral is given by
${\cal O}=\int \psi_1^*({\bf p};{\bf P}_c, {\bf p}_n)\psi_2({\bf p}
;{\bf P}_c, {\bf p}_n)d{\bf p} = \exp(-
{\bf p}_n^2/8\sigma_n^2)$.  It does not depend on ${\bf P}_c$.
Thus, if ${\bf p}_n =0$ the final wave function, $\Psi$, of the pair is 100\% an $S=0$ state,
i.e., $\Psi=\Psi^{(+)}$. For ${\bf p}_n \neq 0$, the singlet and triplet states can both
contribute to $\Psi$.

The probabilities in eq. \ref{CU} can be written as
\begin{equation}
P\left({\bf p}_1; {\bf p}_2 \right)=\int {d\sigma\over d{\bf p}_n d{\bf P}_c} \
\left\{
\left|
\Psi^{(+)}\left(  {\bf p}_1,{\bf p}_2;{\bf P}_c, {\bf p}_n\right) \right|^2
+ {\cal M}
\left|
\Psi^{(-)}\left( {\bf p}_1,{\bf p}_2,;{\bf P}_c, {\bf p}_n\right)
\right|^2  d{\bf P}_c d{\bf p}_n \right\}
\label{PM}
\end{equation}
where ${\cal M}$ is the mixing parameter, determining the relative contribution of the
triplet state, and
$d\sigma/d{\bf P}_cd{\bf p}_n$ is the differential cross section for the process.
The experiments in peripheral collisions with unstable nuclei show that
the momentum distributions are only slightly shifted from ${\bf P}_c=0$ and that the
shift is much smaller than the width of the recoil momentum distribution.
Thus, the correlation function is rather independent of the details of
$d\sigma/d{\bf P}_cd{\bf p}_n$. However, it is important to study the dependence
of the correlation function on ${\bf p}_n$. We thus replace ${\bf P}_c=0$ in the
formulas above and perform the integrals to obtain the corelation function
$R(\Delta p; p_n)$. The result can be obtained analytically. For pure singlet (+), or
pure triplet (-), states one gets
\begin{equation}
R^{(\pm)}={2\left[ g(x)\pm 1 \right] \left[1\pm{\cal O}^2\right] \over
1+2{\cal O}^4+g(x){\cal O}^2\pm 8h(x){\cal O}^{5/2}}-1,
\end{equation}
where $g=\sinh(x)/x$, $h=\sinh(x/2)/x$, and $x=p_n\Delta p/2\sigma^2$.

The results for
$R^{(+)}$ (upper figure) and $R^{(-)}$ (lower figure) are shown in figure 1, as a function
of the variables $\Delta p/\sigma_n$ and $p_n/\sigma_n$.
One sees that the properties of the correlation functions in the singlet
and triplet states are completely different. When the differences of the
average momenta of the nucleons is small $\Delta p \ll \sigma_n$, the correlation
function is negative only for the triplet state. It is -1 at $\Delta p =0$ for
the triplet state, whereas it is close to zero for the singlet state. While for
the former case the correlation function crosses zero at two points, it does not
have a null point for the singlet case.

In figure 2 we show the case for which the final state is an admixture of triplet
and singlet states, as a function of the mixing parameter ${\cal M}$, which appears
in eq. \ref{PM}. The calculation also leads to closed analytical expressions.
The dotted, dashed, and solid curves are obtained for ${\cal M} = 0.1$, 0.5, and 0.9,
respectively. As one increases $\cal M$ the fingerprint of triplet states in the final
wavefunction becomes more and more visible. This means that measurements of pair correlation
functions should be able to discriminate the triplet and singlet configurations in
the final wavefunction.

The traditional idea of diproton radioactivity is due to the pairing effect.
Two protons form a quasiparticle (diproton) under the Coulomb barrier and this
facilitates penetration. In a more formal description, one has a system with two valence
protons in  the same shell and coupled to $J^\pi = 0^+$.
The method described above is directly applicable to determine the spin mixing
of final states in
low-energy two-proton nuclear decay for
$0^+\longrightarrow 0^+$ transitions. In this case the final spin wave function of the
pair equals that of the initial wave function. In particular, when singlet states are
identified,
spin-spin  coincidence experiments will generate dichotomic outcomes for
each single measurement. These can be tested using Bell's inequalities for
spin-spin correlations along different spin-axes. Since not all the particles can be
measured, one would have to use a softer version of the Bell's inequalities, as developed by Clauser
and Horne \cite{CH74}.
In the case of peripheral nuclear reactions this is not so simple. But, the entanglement of the
final wavefunction
will also have noticeable consequences, as we show next.

Let us assume for simplicity the emission of a nucleon pair in a
reaction involving halo nuclei. For weakly bound nucleons one can
assume that the reaction suddenly detaches the pair from the core
and that their intrinsic wavefunction is likely to be an entangled
state. Their wavefunction can be written in the form $\exp(i{\bf
p}_1.{\bf r}_1)\exp(i{\bf p}_2.{\bf r}_2)\pm \exp(i{\bf p}_2.{\bf
r}_1)\exp(i{\bf p}_1.{\bf r}_2)$. This wavefunction leads to
destructive, or constructive interferences. The cross section for
the reaction process will be given by
\begin{equation}
\sigma({\bf P,q}) = f({\bf P,q},R) \left[ 1\pm \cos\left( qR\right)\right] .
\label{sigma}
\end{equation}
where ${\bf q}={\bf p}_1 -{\bf p}_2$, ${\bf P}$ is the recoil momentum of the center-of-mass of
the pair and
$R$ is a parameter which describes the initial spatial localization of the nucleon pair.
All other features
of the reaction mechanism are included in the function $f({\bf P,q},R)$. Thus, for
$qR\ll 1$ one should be
able to see a destructive interference for triplet final states and
constructive interference for singlet final states.
In fact, this was clearly seen in an experiment dedicated to study the existence of the
$^5$H nucleus \cite{Kor01}. The transfer reaction $^1$H($^6$He,$^2$He)$^5$H was studied by
detecting two protons emitted from the decay of $^2$He. The energy correlation of the two
protons was measured, as shown in figure \ref{fig3}. If we apply eq. \ref{sigma} to describe
the relative energy distribution of the two protons, we get
\begin{equation}
\sigma (E_{pp}) \propto \left\{
\begin{array}{c}
\sqrt{E_{pp}} \left[ 1 + \cos \left( \sqrt{m_N E_{pp}/\hbar^2}R\right) \right]\ , \ \ \ \ \ \
{\rm singlet}
\\
\sqrt{E_{pp}} \left[ 1 - \cos \left( \sqrt{m_N E_{pp}/\hbar^2}R\right) \right]\ , \ \ \ \ \ \
{\rm triplet}
\\
\sqrt{E_{pp}}\ , \ \ \ \ \ \ \ {\rm uncorrelated}
\end{array}
\right.
\label{sepp}
\end{equation}

In figure \ref{fig3} we show the result of eqs. \ref{sepp} together with the experimental
data of ref. \cite{Kor01}. The solid, dashed, and dotted curves correspond to singlet, triplet, and
uncorrelated states of the proton pair, respectively.
We used $R=4.6$ fm as the size of the source region where the protons originate.
The agreement with the experiment is quite poor, especially for large relative energy of
the pair. This is expected since we have neglected final state interactions (e.g., Coulomb
interaction). In ref. \cite{Kor01} this spectrum was described theoretically by using the
Landau-Smorodinskii's effective range approximation \cite{ST97} for the protons with a
scattering length of $a_{pp} = -7.806$ fm. However, it is important to notice that the most
important ingredient is the assumption of a singlet or of a triplet state for the protons.
As one easily sees from figure \ref{fig3} the spectrum is completely different in each
situation. When no entanglement exists, there is no interference and the spectrum also changes
dramatically.

The examples discussed above involve nucleon pairs interacting with one or more particles.
Entanglement measures for general multiparticle systems are still under dispute \cite{VPR97}. However, if
the state of the whole system is a pure state, and the full system is being regarded as divided
into two subsystems, a convenient measure of entanglement is
\begin{equation}
P_n=Tr\left\{\rho_n^2(t)\right\} = Tr_n\left\{ \left[ Tr_c\left(
|\Psi><\Psi|\right)\right]^2\right\},
 \label{measure}
\end{equation}
where $\rho_n$ is the reduced density operator of the nucleon pair subsystem, $N$, and
$|\Psi>$ is the wavefunction of the total system. Under these conditions, it can be
shown \cite{PTT97} that this quantity yields the same value no matter for which
subsystem it is
calculated, i.e., $P_c=P_n\equiv P$ (the traces in eq. \ref{measure} can
be permuted). $P$ is called the ``purity" since it takes on its
maximum value 1 if the subsystem is in a pure state. The more $P$ deviates from 1 the
larger is the error that would occur in treating the system as an entangled state.
To illustrate this point, let us consider the fragmentation of $^{11}$Li projectiles,
as obtained in the experiment \cite{Sim99}.

The analysis of experiment \cite{Sim99} was based on Hansen's wounded wavefunction model
\cite{Han96}. This method assumes
that the collision is so fast that a piece of the initial wavefunction is scraped-off
in the collision. This allows to calculate the frozen momentum configuration of the initial
state by a Fourier transform of this wavefunction. One therefore assumes a 100\%
entanglement of the remaining constituents of the nucleus. In the experiment  \cite{Sim99} this
was used in the case of a knock out reaction, where a neutron was removed from $^{11}$Li.
The measurement of the momentum distributions of the neutron and the $^9$Li fragment
was used to deduce the properties of $^{10}$Li.
The nucleus $^{11}$Li is usually called a Borromean nucleus, meaning that both the
n-n, as well as the $^{10}$Li, subsystems are not bound. We now show that it is
exactly this property which is responsible for the successful
interpretation of the experiment  \cite{Sim99}.
We use eq.
\ref{measure} for the subsystems $^{10}$Li (c) and n-n (N).

We use the results of
ref. \cite{JJH90}, where it was shown that the bound-state
properties of $^{11}$Li can be obtained  with a three-body
model by using phenomenological potentials of the form
\begin{equation}
V_{nn}=S_n e^{-\rho^2/b_n^2}, \ \ \ \ V_{n9}=S_1 e^{-\lambda^2/c_1^2}+S_2 e^{-\lambda^2/c_2^2},
\end{equation}
where
\begin{equation}
\mbox{\boldmath$\rho$}={\bf r}_n - {\bf r}_{n'},\ \ \ \ \  \mbox{\boldmath$\lambda$}=
{1\over 2}\left( {\bf r}_n + {\bf r}_{n'}\right) -{\bf r}_9,
\end{equation}
are the Jacobian variables. The following potential parameters were used:
(a) $S_n=-31$ MeV, $b_n=1.8$ fm, to reproduce the low-energy scattering length and
effective range of the n-n system. (b) The set of values $S_1=-7.0$ MeV, $c_1 = 2.4$ fm,
$S_2=-1.0$ MeV, $c_2=3.0$ fm, reproduce the size and the binding of $^{11}$Li
($\sim 0.3$ MeV) quite reasonably.

In eq. \ref{measure} one should insert all bound states of the
subsystems to obtain the probability that a measurement of the Hansen's wavefunction
would give an entangled $^{10}$Li nucleus.
However, there are no bound states in these systems. In fact,
we notice that the potentials presented above barely bind the n-n and the $^9$Li subsystems.
They yield the (negative) scattering lengths, $a_{nn}=-17.4$ fm and
$a_{n9}=-10.66$ fm, respectively. These are
large values, reflecting almost bound states, with $E\simeq 0$. Thus, one can estimate
the value of the purity in eq. \ref{measure} by replacing  their respective wavefunction by a
constant value within $^{11}$Li. This yields the particularly simple expression
\begin{equation}
P\simeq \left({1\over 4\pi <\lambda>^3/3}\right)^{1/2}
\left({1\over 4\pi <\rho>^3/3}\right)^{1/2}
\int d\mbox{\boldmath $\rho$}
\left\{\int d\mbox{\boldmath $\lambda$}\Psi_{^{11}{\rm Li}}
\left(\mbox{\boldmath $\rho$},\mbox{\boldmath $\lambda$}\right) \right\}^2,
\label{Pf}
\end{equation}
where $\Psi_{^{11}{\rm Li}}$ is the $^{11}$Li wavefunction in terms of the Jacobian variables
and $<\rho>=3.3$ fm and $<\lambda>=3.1$ fm are the most probable values of the
Jacobian coordinates.
The $^{11}$Li wavefunction in eq. \ref{Pf} is obtained from a variational calculation, as in
\cite{JJH90}. We get $P\simeq 0.8$ which is very close to unity, thus proving our assertion.

In conclusion, we have shown that entangled states in  nuclear decay and in reactions with
exotic nuclei is another route to study important problems of basic quantum
mechanics interest, as the Einstein-Podolski-Rosen paradox. These studies would be
complementary to
others performed in atomic physics. We have also shown that, due to the entanglement
of final states, simple correlation functions
yield compelling evidence of the spin and momentum states of the nuclear
subsystems.
The forthcoming RIA (Rare Isotope Accelerator) facility will be an ideal laboratory to perform
these studies \cite{RIA}.

\section*{Acknowledgements}
This research was supported in part by the U.S. National Science
Foundation under Grants No. PHY-007091 and PHY-00-70818.

\newpage

%%%%%%%%%%%%%%%%%%%%%%%%%%%%%%%%%%%%%%%%%%%%%%%%%%%%%%%%%%
\begin{figure}[t]
\centerline{\psfig{file=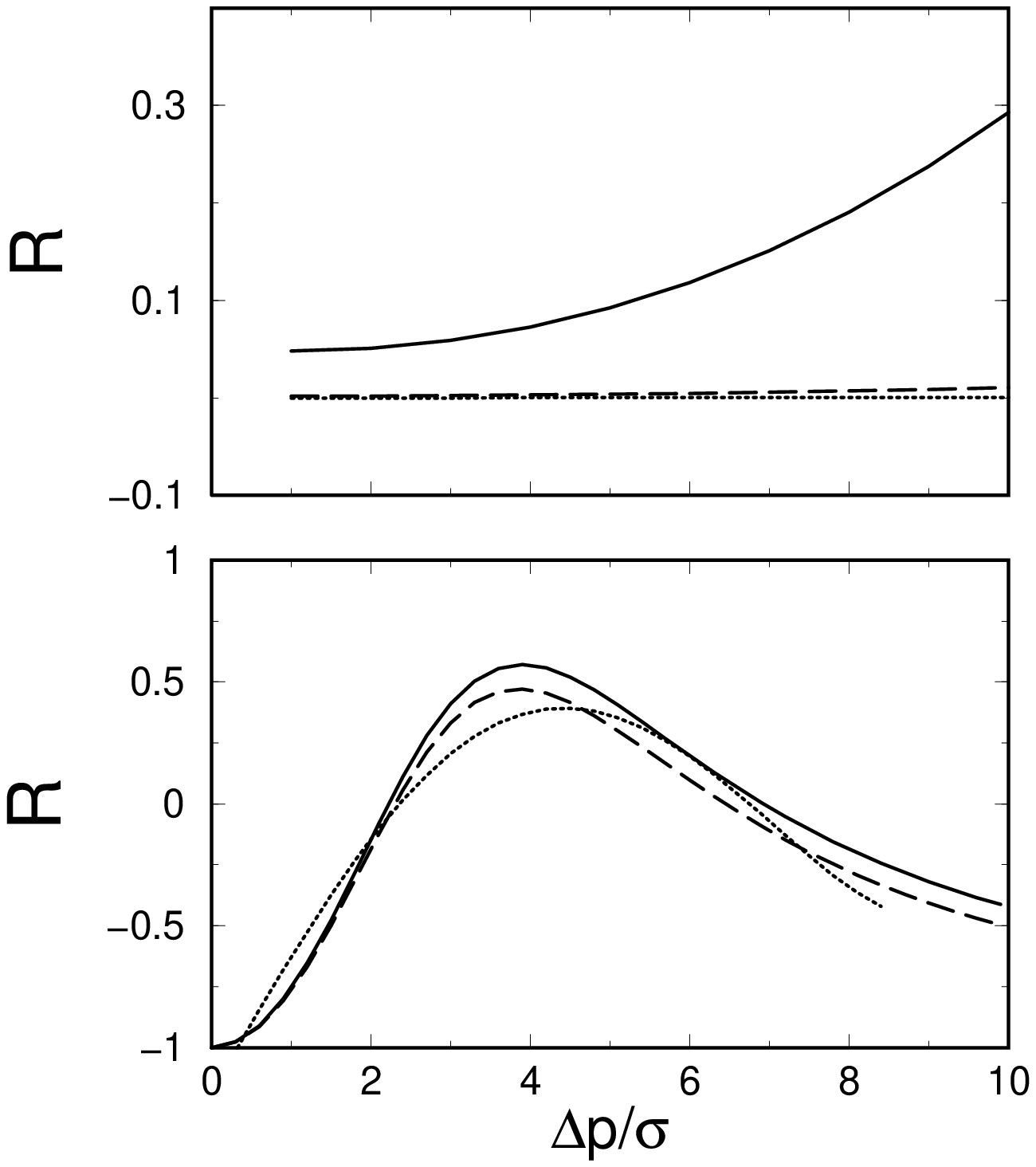,width=100mm}}
\caption{
Correlation functions for different values of the average
momentum, $p_n$. The upper (lower) figure is for the singlet
(triplet) state.
The dotted, dashed, and solid lines correspond to
$p_n/\sigma_n = 0.01$, 0.1, and 0.5, respectively.
} \label{fig1}
\end{figure}
%%%%%%%%%%%%%%%%%%%%%%%%%%%%%%%%%%%%%%%%%%%%%%%%%%%%%%%%%%%%%%

\newpage

%%%%%%%%%%%%%%%%%%%%%%%%%%%%%%%%%%%%%%%%%%%%%%%%%%%%%%%%%%
\begin{figure}[t]
\centerline{\psfig{file=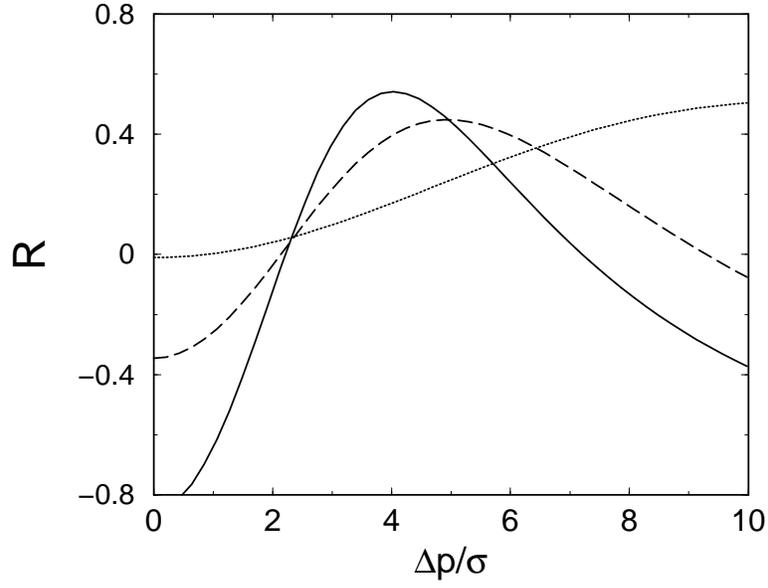,width=100mm}}
\caption{
Correlation functions for $p_n/\sigma_n=0.5$ and for different admixtures of singlet
and triplet states. The dotted, dashed, and solid lines correspond to
${\cal M} = 0.1$, 0.5, and 0.9, respectively. ${\cal M}$ is the absolute
contribution of the triplet state.
} \label{fig2}
\end{figure}
%%%%%%%%%%%%%%%%%%%%%%%%%%%%%%%%%%%%%%%%%%%%%%%%%%%%%%%%%%%%%%

\newpage

%%%%%%%%%%%%%%%%%%%%%%%%%%%%%%%%%%%%%%%%%%%%%%%%%%%%%%%%%%
\begin{figure}[t]
\centerline{\psfig{file=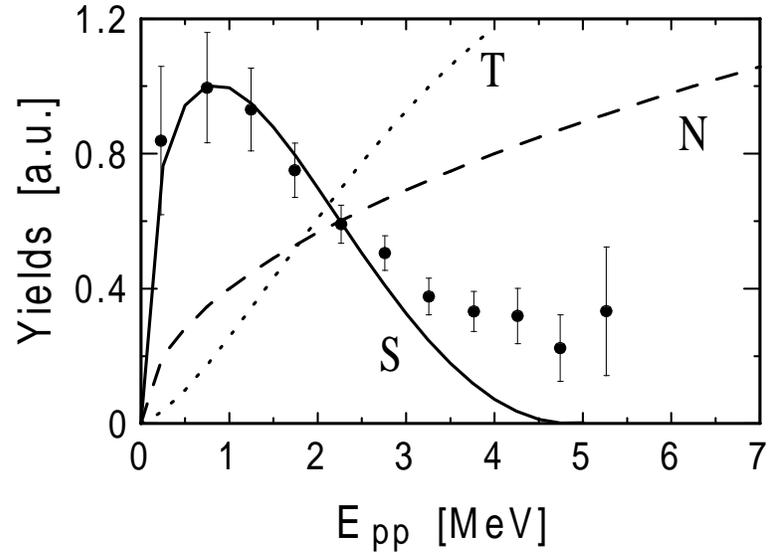,width=100mm}}
\caption{
Relative energy distribution of two protons from the reaction
p($^6$He, ppt). Calculations are for
triplet (T) and singlet (S) states. The dotted curve assumes no correlation
in the pair wavefunction.
} \label{fig3}
\end{figure}
%%%%%%%%%%%%%%%%%%%%%%%%%%%%%%%%%%%%%%%%%%%%%%%%%%%%%%%%%%%%%%

$\,\,\,\,\,\,\,\,\,\,\,\,\,\,\,\,\,\,\,\,\,$

\end{document}